\documentclass[letterpaper, 10 pt, conference]{ieeeconf}  

\IEEEoverridecommandlockouts                              

\usepackage{graphicx} 
\usepackage{amsmath} 
\usepackage{algorithm}
\usepackage{algpseudocode}
\usepackage{tabularx}
\usepackage{booktabs}
\usepackage{tfrupee}
\usepackage{inputenc}
\usepackage{booktabs}
\usepackage{multirow}
\usepackage{graphicx}
\usepackage{soul}
\usepackage{xcolor}
\usepackage{url}
\usepackage{subcaption}

\title{\LARGE \bf
Towards Effective Deep Neural Network Approach for Multi-Trial P300-based Character Recognition in Brain-Computer Interfaces}

\author{Praveen Kumar Shukla$^{1}$, Hubert Cecotti$^{2}$, and  Yogesh Kumar Meena$^{3}$
\thanks{$^{1,3}$ Praveen Kumar Shukla and Yogesh Kumar Meena are with Human-AI Interaction (HAIx) Lab, IIT Gandhinagar, India
        {\tt\small yk.meena@iitgn.ac.in}}%
\thanks{$^{2}$ Hubert Cecotti  is with the Department of Computer Science, California State University, Fresno, USA }
}

\begin{document}

\maketitle
\thispagestyle{empty}
\pagestyle{empty}

\begin{abstract}

Brain-computer interfaces (BCIs) enable direct interaction between users and computers by decoding brain signals. This study addresses the challenges of detecting P300 event-related potentials in electroencephalograms (EEGs) and integrating these P300 responses for character spelling, particularly within oddball paradigms characterized by uneven P300 distribution, low target probability, and poor signal-to-noise ratio (SNR). This work proposes a weighted ensemble spatio-sequential convolutional neural network (WE-SPSQ-CNN) to improve classification accuracy and SNR by mitigating signal variability for character identification. We evaluated the proposed WE-SPSQ-CNN on dataset II from the BCI Competition III, achieving P300 classification accuracies of 69.7\% for subject A and 79.9\% for subject B across fifteen epochs. For character recognition, the model achieved average accuracies of 76.5\%, 87.5\%, and 94.5\% with five, ten, and fifteen repetitions, respectively.  Our proposed model outperformed state-of-the-art models in the five-repetition and delivered comparable performance in the ten and fifteen repetitions.

\end{abstract}

\section{Introduction}
\label{sec:introduction}

Patients suffering from severe muscular disorders or neuropathy typically find it difficult or impossible to interact with the outside world. Brain-computer interface (BCI) technology helps these patients communicate with the outside world~\cite{hochberg2012reach}. BCI system interprets brain activity and produces a control command for an external device ~\cite{birbaumer2007brain}. According to techniques for recording brain signals, BCI systems can be divided into two main categories: non-invasive and invasive. This work considers the non-invasive BCI, which is less expensive, safer, and has a lower signal-to-noise ratio (SNR). 

Electroencephalography (EEG) is the key sensing technology in non-invasive BCI systems. Several types of brain responses, like P300~\cite{shukla2020brain}, steady-state evoked potential (SSVEP)~\cite{hwang2012development}, and hybrid~\cite{{7318410}, {6999180}, {su2011hybrid}}, are used for BCI systems. These BCI systems offer multiple applications, such as keyboard control \cite{krusienski2011control}, home appliances~\cite{shukla2021performance}, and character recognition \cite{chaurasiya2016binary}. In this work, we consider the P300-based BCI system for character recognition. The signal is identified as P300 because it shows a notable peak approximately 300 milliseconds after the stimulus.

Numerous methods have emerged in recent years for extracting features and classification in the context of P300 and Non-P300, particularly in character recognition. For instance, Rakotomamonjy et al.~\cite{4454051} achieved a 96.5\% classification accuracy in 15 repetitions (epochs/trials) by employing a recursive channel elimination method and an ensemble of support vector machine (ESVM) classification approach. Salvaris et al.~\cite{5109302} proposed the utilization of wavelet for feature extraction and an ensemble of fisher’s linear discriminant (FLD) for P300 classification. Cecotti et al.~\cite{5492691} recommended a combination of four single and three multiple classifiers based on convolutional neural network (CNN), attaining the highest accuracy of 95.5\%. Bhatnagar et al.~\cite{7449163} introduced the ensemble of SVM as a classification method for single-trial detection, achieving an accuracy of 92.5\%. Kundu et al.~\cite{8493903} suggested an ensemble of SVM specifically for P300 classification. Furthermore, principal component analysis (PCA) was applied for data reduction and feature extraction, while ESVM served as the classification method for character recognition~\cite{doi:10.1080/03772063.2017.1355271}. In another work~\cite{9031908}, two CNN networks with distinct kernel dimensions were employed for character detection. Despite these advancements, there still needs to be a gap in achieving high accuracy, particularly in scenarios involving fewer repetitions, hence increasing the information transfer rate. 



To address the inherent imbalance between P300 and Non-P300 signals~\cite{{8932567},{kundu2019p300},{7582466},{9427074},{1454155},{9516980}}, our approach divides Non-P300 signals into five equal portions and replicates the P300 signals four times to create five copies. Each P300 copy is paired with a portion of Non-P300 signals, generating five balanced subsets. To improve the low SNR in EEG data, we apply signal averaging. We then use an ensemble of classifiers to reduce signal variability and enhance SNR. Leveraging the strengths of CNNs for binary classification, our method employs a CNN ensemble to achieve higher accuracy in P300 classification and character recognition.

The subsequent sections of the paper are structured as follows: Section~\ref{sec:methods} details the methodology, encompassing components such as the proposed architecture, paradigm design, dataset description, data pre-processing, and the proposed model. Section~\ref{sec:Results} presents the results. Finally, the key findings are summarized in Section~\ref{sec:discussion}. 



\begin{figure}[t]
    \centering
    \includegraphics[width=0.5\textwidth]{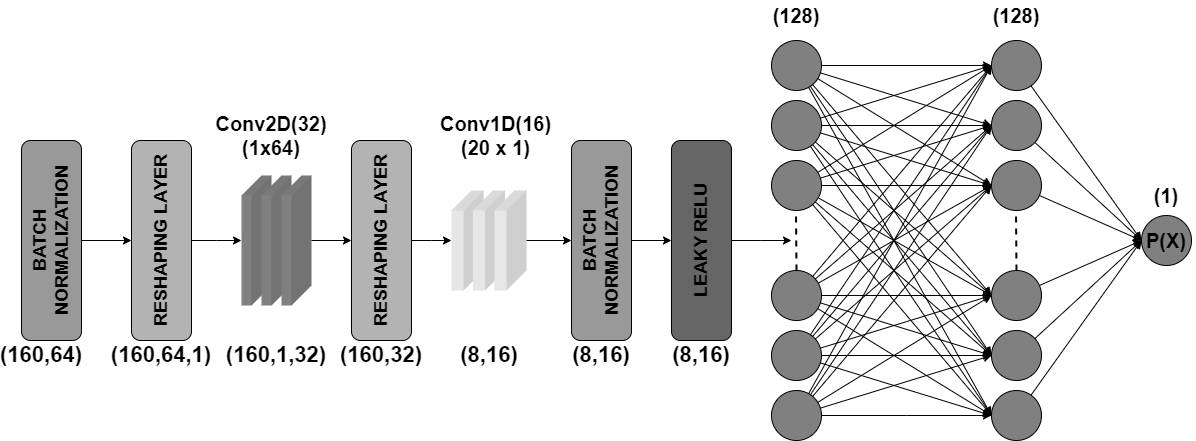}
    \caption{The architecture of proposed spatio-sequential CNN model for P300 and non-300 classification.}
    \label{fig:figurelabel1}
\end{figure}


\section{ Methods}
\label{sec:methods}

\subsection{Spatio-sequential CNN}

The base classifier used for the proposed weighted ensemble is called spatio-sequential CNN (SPSQ-CNN). Fig.~\ref{fig:figurelabel1} presents the full architecture of the base classifier, with a detailed layer-by-layer description in Table~\ref{tab:1}. The SPSQ-CNN integrates a 2-dimensional convolution kernel for spatial feature extraction and a 1-dimensional convolution kernel for temporal feature extraction. Two batch normalization layers have been employed to normalize the input signal and prevent the internal covariate shift which is the change in the distribution of network activations due to the change in network parameters during training. This makes the training process much faster and increases the generalizability of the model. The second batch normalization layer is followed by a leaky rectified linear unit (ReLU). This activation prevents the problem of dying ReLU and promotes smooth gradient flow through the network hence enhancing the learning capability of the model. The model's top is fitted with 2 fully connected layers each having a hyperbolic tangent (tanh) activation function and dropout rate (p) of 0.8 each. The output layer of the classifier gives the probability of the input being a P300 signal using the sigmoid activation function.  The model has been constructed and trained using the keras API with tensorFlow backend on a T4 GPU on a google colab environment. The batch size for the gradient descent was set to 32 batches and the optimizer used was adam with a learning rate of 0.001. The reason for using adam over others is its ability to provide fast convergence and hence completion of training in less number of epochs. For the learning process, binary cross-entropy loss was used.




\begin{table}[]
\caption{The proposed spatio-sequential CNN model consists of layers with kernel sizes, output sizes, and number of parameters for corresponding layers.}
\label{tab:1}
\begin{tabular}{@{}lccc@{}}
\toprule
Layer Type & Kernel Size & Output & Parameters \\ \midrule
Batch Normalization 1 & --- & 160,64 & 256 \\
Reshape 1 & --- & 160,64,1 & 0 \\
2D Convolution & 1 x 64 & 160,1,32 & 2,080 \\
Reshape 2 & --- & 160,32 & 0 \\
1D Convolution & 20 x 1 & 8,16 & 10,256 \\
Batch Normalization 2 & --- & 8,16 & 64 \\
Leaky ReLU Activation & --- & 8,16 & 0 \\
Fully Connected 1 & 128 & 128 & 16,512 \\
Fully Connected 2 & 128 & 128 & 16,512 \\
Fully Connected 3 (output) & 1 & 1 & 129 \\ \midrule
\multicolumn{3}{l}{Total Parameters} & 45,809 \\ \bottomrule
\end{tabular}
\end{table}

\subsection{Weighted ensemble of SPSQ-CNNs}

The five base classifiers show highly variable performance due to the differences in the data quality and distribution of the training subset. A weighted ensemble is proposed, assigning weights to each model based on a metric. The chosen metric is the validation accuracy on the respective dataset subset. We calculate weights using the Eq.~\ref{eq1}. 

\begin{equation}
W_k=\frac{T_k}{\sum_{i=1}^n T_i}
\label{eq1}
\end{equation}

Where $T_i=TP_i+TN_i$: Total true prediction by the i\textsuperscript{th} classifier, $TP_i$ is the true positive detected by the i\textsuperscript{th} classifier, $TN_i$ is the true negatives detected by the i\textsuperscript{th} classifier.

\subsection{P300 speller}

\begin{figure*}[t]
\centering
 \includegraphics[width=0.94\textwidth]{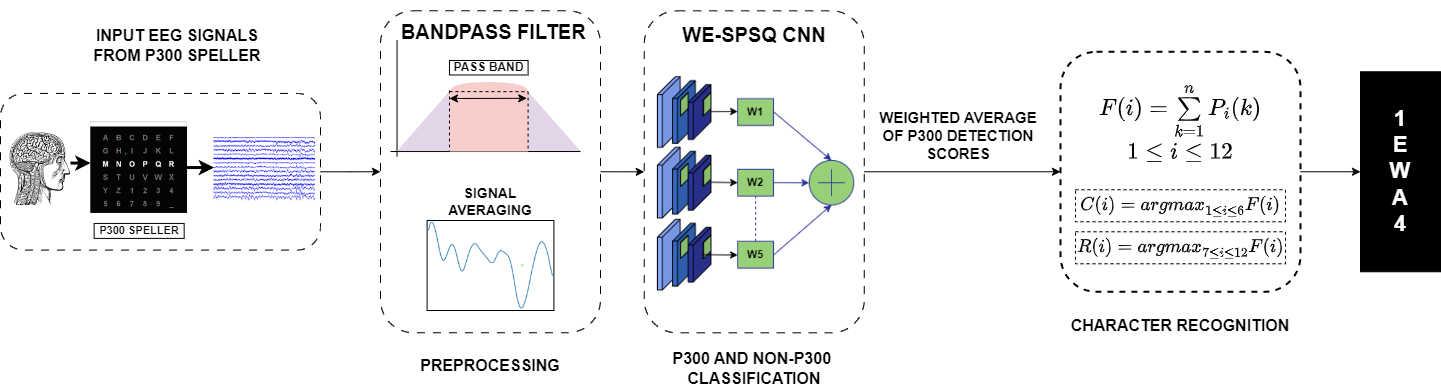}
\caption{The schematic block diagram outlines our proposed framework for character identification in RC-based P300 Speller. It employs a band-pass filter and signal averaging for data prepossessing, with P300 classification and character recognition achieved through WE-SPSQ CNN classification.}
\label{fig:figurelabel}
\end{figure*}

Fig.~\ref{fig:figurelabel} depicts a schematic diagram of P300-based character detection. The proposed method is divided into three major phases: data preprocessing, P300 and Non-P300 detection, and character classification. The preprocessing phase uses bandpass filtering, averaging for noise removal, and subsampling followed by a bootstrap-based ensemble of spatio-sequential CNNs for P300 and Non-P300 classification. Finally, using a score-based averaging mechanism, the target character is predicted. 

\begin{figure}[t]
    \centering
    \includegraphics[width=0.24\textwidth]{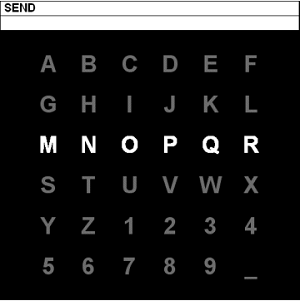}
    \caption{Participants are presented with the target word 'SEND' in the P300-based Row-Column BCI Paradigm~\cite{farwell1988talking}}
    \label{fig:p300}
\end{figure}

Farwell and Donchin~\cite{farwell1988talking} proposed the P300 speller paradigm. The data was collected using the BCI speller paradigm, as shown in Fig.~\ref{fig:p300} with the standard row-column paradigm with a 6 x 6 matrix with 36 symbols (26 letters, 9 digits, and 1 symbol), 12 flashes (6 rows, 6 columns) for the selection of a cell. 


The matrix is organized into six rows and six columns, with both rows and columns randomly flashing during the BCI paradigm. Each flash is intensified for 100 ms, followed by a blank period of 75 ms. A P300 is elicited when the desired symbol appears in the flashed row or column. To capture the P300 wave, a window of 667~ms post-stimulus is considered for each intensification, accounting for the 300 ms latency in the creation of the P300 ERP component. While the paradigm is named after the P300 ERP component, all the EEG signal post-stimulus is included in the classification, which contains other ERP components.

This experiment utilizes the BCI Competition III data set II, specifically designed for the P300 speller paradigm \cite{farwell1988talking}. BCI competition III dataset II is provided by the BCI laboratory of the wadsworth center, NYS department of health. Two participants contribute to the dataset through the experiment. During the task, users were instructed to focus on a desired symbol within a word they intended to communicate. The dataset comprises 85 training symbols and 100 testing symbols for each subject, with participants wearing a headset equipped with 64 electrodes.

We used their training dataset because both individuals have labels. The matrix is organized into six rows and six columns, with both rows and columns randomly flashing during the BCI paradigm. Each flash is intensified for 100 ms, followed by a blank period of 75 ms. Out of 6 rows and 6 columns, one row and one column contain the desired character. One {epoch (trial) is defined as these 12 intensifications. This process is repeated fifteen times per epoch to enhance accuracy, resulting in a total of 12 × 15 = 180 flashing epochs for one symbol. The EEG signals are recorded at a sampling rate of 240 Hz and undergo band-pass filtering with a cut-off frequency range of 0.1 - 60 Hz. The desired symbol appears in the flashed row or column, evoking a P300 signal. There are 15300 samples collected using 64 different recording channels for both subjects. Out of 15300 samples, data has 2550 samples for P300 and 12750 samples for Non-P300
in training 3000 samples for P300 and 15000 samples for Non-P300 in testing.




For a single trial (i.e., 12 flashes), the number of Non-P300 signals detected are 10 and the number of P300 signals detected are 2. This shows that the number of Non-P300 signals are five times more than the P300 signals. In addressing this inherent imbalance between P300 and Non-P300 signals,
our approach strategically divides Non-P300 signals into five equal sized subsets and the P300 signals are cloned 4 more times giving us 5 clones of the P300 signals. Each P300 signal clone is combined with a subset of the Non-P300 signal ensuring a balanced distribution of both types of signals per subset (2,550 signals of each type in each subset)}

\subsubsection{Character classification}

For predicting the target character, the weighted score of the output prediction of P300 and Non-P300 is taken, and a score is assigned to them. The target rows and columns are detected using the Eq.~\ref{eq2}. 

\begin{equation}
\mathrm F(i)=\frac{1}{j}\frac{1}{k}\sum_{j=1}^{n}\sum_{k=1}^{5}\mathrm{W}_{k}\ast \mathrm{p}_{k}^{}
\label{eq2}
\end{equation}

\begin{equation}
C_i=\operatorname{argmax}_{1<i<6} F(i)
\label{eq}
\end{equation}
\begin{equation}
R_i=\operatorname{argmax}_{7<i<12} F(i)
\label{eq}
\end{equation}

\begin{table*}[]
\centering
\caption{Result for character recognition simulation for both subject A and B.}
\label{tab:simu}
\begin{tabular}{@{}c|ccccccccccccccc@{}}
\toprule
\multirow{2}{*}{Subject} & \multicolumn{15}{c}{Number of repetitions (epochs/trials)} \\
 & 1 & 2 & 3 & 4 & 5 & 6 & 7 & 8 & 9 & 10 & 11 & 12 & 13 & 14 & 15 \\ \midrule
A & 49.1 & 77.1 & 90.7 & 95.9 & 98.1 & 99 & 99.6 & 100 & 99.9 & 100 & 100 & 100 & 100 & 100 & 100 \\
B & 48.2 & 76.9 & 89.2 & 95.0 & 97.9 & 99.1 & 99.6 & 99.9 & 99.9 & 99.9 & 99.9 & 100 & 100 & 100 & 100 \\ \bottomrule
\end{tabular}
\end{table*}


\begin{table*}[]
\centering
\caption{Result for P300 classification for subject A.}
\label{tab:p300sA}
\begin{tabular}{@{}lcccccccc@{}}
\toprule
\multicolumn{1}{c}{Method} & TP & TN & FN & FP & Pres. & Rec. & Acc. & F1 \\ \midrule
SPSQ CNN 1 & 1053 & 9081 & 1947 & 5919 & 0.351 & 0.151 & 0.563 & 0.210 \\
SPSQ CNN 2 & 374 & 12906 & 2626 & 2094 & 0.124 & 0.151 & 0.737 & 0.136 \\
SPSQ CNN 3 & 840 & 10489 & 2160 & 4511 & 0.28 & 0.157 & 0.629 & 0.199 \\
SPSQ CNN 4 & 2327 & 3118 & 673 & 11882 & 0.77 & 0.163 & 0.302 & 0.269 \\
SPSQ CNN 5 & 259 & 13624 & 2741 & 1376 & 0.086 & 0.158 & 0.771 & 0.1113 \\
WE-SPSQ CNN & --- & --- & --- & --- & 0.177 & 0.150 & \textbf{0.697} & 0.162 \\ \bottomrule
\end{tabular}
\end{table*}


\begin{table*}[]
\centering
\caption{Result for P300 classification for subject B.}
\label{tab:p300sB}
\begin{tabular}{@{}lllllllll@{}}
\toprule
Method & \multicolumn{1}{c}{TP} & \multicolumn{1}{c}{TN} & \multicolumn{1}{c}{FN} & \multicolumn{1}{c}{FP} & \multicolumn{1}{c}{Pres.} & \multicolumn{1}{c}{Rec.} & \multicolumn{1}{c}{Acc.} & \multicolumn{1}{c}{F1} \\ \midrule
SPSQ CNN 1 & 141 & 14391 & 2859 & 609 & 0.047 & 0.188 & 0.80 & 0.0752 \\
SPSQ CNN 2 & 495 & 12489 & 2505 & 2511 & 0.165 & 0.164 & 0.721 & 0.164 \\
SPSQ CNN 3 & 319 & 13256 & 2681 & 1744 & 0.106 & 0.154 & 0.754 & 0.125 \\
SPSQ CNN 4 & 251 & 13425 & 2749 & 1575 & 0.0836 & 0.137 & 0.759 & 0.103 \\
SPSQ CNN 5 & 341 & 13124 & 1876 & 2659 & 0.113 & 0.154 & 0.748 & 0.130 \\
WE-SPSQ CNN & --- & --- & --- & --- & 0.046 & 0.154 & \textbf{0.799} & 0.0708 \\ \bottomrule
\end{tabular}
\end{table*}


\subsubsection{Signal preprocessing}
Where $F(i)$ represents the final score for the i\textsuperscript{th} row and column, j represents the number of trials, k represents the number of classifiers,
${W}_{k}$ represent weight of the i\textsuperscript{th} classifier, ${P}_{k}$ represent the prediction mode by the k\textsuperscript{th} classifier. $C_{i}$ and $R_{i}$ represent the predicted column and row, respectively. The row and column intersection ($R_i$, $C_i$) tells the character predicted.

The P300 information in the signal is captured within a 667~ms data window following the stimulus. Categorization is performed for each channel using a 0-667 ms window with 160 samples. A Type I 4th order bandpass chebyshev filter, with a cut-off frequency ranging from 0.1 to 10 Hz, filters the signal from each channel. The filtered signal is then decimated every 14th sample from the 160-sample window. Subsequently, the decimated samples from all 64 channels are concatenated to form a vector. As a result, each subject in the training set is represented by a vector of dimensions 15300 $\times$ 896 [15300 (12 rows/columns × 15 repetitions $\times$ 85 characters), dimension of 896 (14 Samples $\times$ 64 channels)].

\subsection{Simulations}

We consider area under the receiver operating characteristics (ROC) curve (AUC) as the main classifier performance for single-trial detection. Using the AUC, we can define $d'$ which is calculated as the difference between the z-score of the hit rate (correctly identifying the signal) and the z-score of the false alarm rate (incorrectly identifying noise as signal).
\begin{eqnarray}
d' & = & Z(hit\ rate)-Z(false \ alarm \ rate)
\end{eqnarray}
where $Z$ is the inverse of the cumulative distribution function of the standard normal distribution.

\begin{eqnarray}
d'  =  \sqrt{2} * norminv(auc)
\end{eqnarray}

where $norminv(p)$ returns the inverse of the standard normal cumulative distribution function (cdf). It is then possible to generate decisions based on the AUC using random numbers coming from a standard normal distribution. Based on the AUC, it is therefore possible to simulate the accumulated score for rows and columns and assess the performance for any number of repetitions. The accuracy using simulated decisions using $d'$ is given in Table~\ref{tab:simu}.

\begin{table*}[]
\centering
\caption{Result for character recognition for both subject A and B.}
\label{Table V}
\begin{tabular}{@{}c|c|ccccccccccccccc@{}}
\toprule
\multirow{2}{*}{Method} & \multirow{2}{*}{Subject} & \multicolumn{15}{c}{Number of repetitions (epochs/trials)} \\
 &  & 1 & 2 & 3 & 4 & 5 & 6 & 7 & 8 & 9 & 10 & 11 & 12 & 13 & 14 & 15 \\ \midrule
\multirow{4}{*}{WE-SPSQ CNN} & A & 19 & 29 & 56 & 67 & 75 & 70 & 76 & 77 & 83 & 86 & 88 & 91 & 94 & 93 & 98 \\
 & B & 43 & 59 & 66 & 77 & 78 & 83 & 89 & 89 & 89 & 93 & 92 & 92 & 91 & 92 & 91 \\
 & Mean & 31 & 44 & 61 & 72 & 76.5 & 76.5 & 82.5 & 83 & 86 & 87.5 & 90 & 91.5 & 92.5 & 92.5 & 94.5 \\
 & Std & 16.9 & 21.2 & 7.0 & 7.0 & 2.1 & 9.1 & 9.1 & 8.4 & 4.2 & 4.9 & 2.8 & 0.7 & 2.1 & 0.7 & 4.9 \\ \bottomrule
\end{tabular}
\end{table*}


\begin{table*}[]
\centering
 \caption{State-of-the-art vs. proposed model performance in 5,10,15 repetitions(Epochs)}
\label{Table IX}
\begin{tabular}{@{}lllll@{}}
\toprule
\multicolumn{1}{c}{\multirow{2}{*}{Author}} & \multicolumn{1}{c}{\multirow{2}{*}{Model}} & \multicolumn{3}{c}{Accuracy} \\
\multicolumn{1}{c}{} & \multicolumn{1}{c}{} & \multicolumn{1}{c}{5 epochs} & \multicolumn{1}{c}{10 epochs} & \multicolumn{1}{c}{15 epochs} \\ \midrule
Rakotomamonjy et al.{}\cite{4454051}{}  & ESVM & 73.5 & 87.0 & 96.5 \\
Salvaris et al.{}\cite{5109302}{} & FLD & 73.5 & 87 & 96.5 \\
Cecotti et al.{}\cite{5492691}{} & CNN-1 & 70 & 88.5 & 94.5 \\
Cecotti et al.\cite{5492691}{} & MCNN-1 & 69 & 87 & 95.5 \\
Bhatnagar et.al \cite {7449163}{}& Ensemble of SVM & 70.5 & 82 & 92.5 \\
Kundu et.al \cite {doi:10.1080/03772063.2017.1355271} & PCA- EWSVM & 72 & 87.5 & 98 \\
Kundu et.al \cite {9031908}{} & Ensemble of CNN & 73.0 & 90.5 & 96.5 \\
Zhang et.al \cite{9516980}{}{}& SVNN & 74.5 & 92 & 98 \\
Wang et.al \cite{10018278}{}& ST-CapsNet & 68 & 88.7 & 98 \\
Our proposed & WE-SPSQ CNN & 76.5 & 87.5 & 94.5 \\ \bottomrule
\end{tabular}
\end{table*}

\begin{table*}[]
\caption{Results of ablation study for proposed model performance.}
\label{Table VII}
\begin{tabular}{@{}c|c|ccccccccccccccc@{}}
\toprule
\multirow{2}{*}{Method} & \multirow{2}{*}{Subject} & \multicolumn{15}{c}{Number of repetitions (epochs/trials)} \\
 &  & 1 & 2 & 3 & 4 & 5 & 6 & 7 & 8 & 9 & 10 & 11 & 12 & 13 & 14 & 15 \\ \midrule
\multirow{4}{*}{SPSQ CNN} & A & 16 & 23 & 30 & 39 & 52 & 52 & 62 & 62 & 70 & 76 & 77 & 80 & 79 & 80 & 84 \\
 & B & 36 & 47 & 54 & 61 & 72 & 73 & 73 & 77 & 81 & 80 & 85 & 86 & 87 & 87 & 88 \\
 & Mean & 26 & 35 & 42 & 50 & 62 & 62.5 & 67.5 & 69.5 & 75.5 & 78 & 81 & 83 & 83 & 83.5 & 86 \\
 & Std & 14.1 & 16.9 & 16.9 & 15.5 & 14.1 & 14.8 & 7.7 & 2.8 & 9.9 & 7.7 & 5.6 & 4.2 & 5.6 & 4.9 & 2.8 \\
\multirow{4}{*}{WE-SPSQ CNN} & A & 19 & 29 & 56 & 67 & 75 & 70 & 76 & 77 & 83 & 86 & 88 & 91 & 94 & 93 & 98 \\
 & B & 43 & 59 & 66 & 77 & 78 & 83 & 89 & 89 & 89 & 93 & 92 & 92 & 91 & 92 & 91 \\
 & Mean & 31 & 44 & 61 & 72 & 76.5 & 76.5 & 82.5 & 83 & 86 & 87.5 & 90 & 91.5 & 92.5 & 92.5 & 94.5 \\
 & Std & 16.9 & 21.2 & 7.0 & 7.0 & 2.1 & 9.1 & 9.1 & 8.4 & 4.2 & 4.9 & 2.8 & 0.7 & 2.1 & 0.7 & 4.9 \\ \bottomrule
\end{tabular}
\end{table*}

\begin{figure}[h]
  \centering
  \begin{tabular}{cc}
\includegraphics[width=.36\textwidth,keepaspectratio]{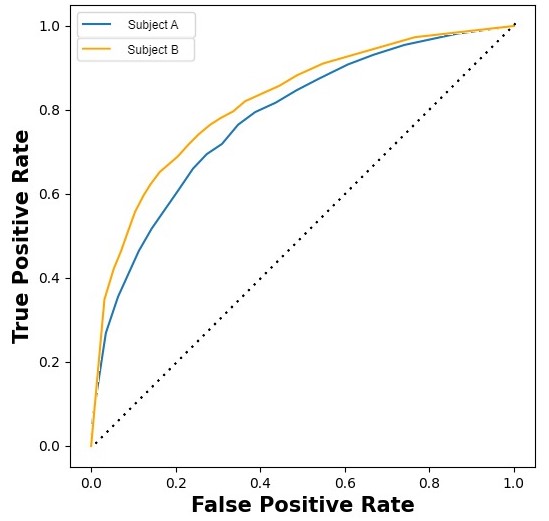} &
   \end{tabular}
\caption{ROC for both subjects A and B.}
  \label{AUC_ROC (2).jpg}
\end{figure}

\section{Results }
\label{sec:Results}

\subsection{P300 speller}



The proposed weighted ensemble consists of five base classifiers. Tables III and IV display the P300 detection performance of each base classifier and the weighted ensemble. As shown, the classifiers tend to predict more true negatives than true positives. This is because the classifiers have been trained on a balanced subset of the training data but the testing data is much larger than the training subset and is highly skewed toward the Non-P300 values. It can be observed from tables~\ref{tab:p300sA} and~\ref{tab:p300sB} that all the classifiers show better performance for subject B as compared to subject A. This can also be observed from the receiver operating characteristics (ROC) curve shown in Fig ~\ref{AUC_ROC (2).jpg}. It shows that for subject B, the proposed ensemble can differentiate better between P300 and Non-P300 signals at different thresholds as compared to signals from subject A as the Area Under the ROC Curve (AUC) for subject B is higher than that for subject A. P300 detection as a binary classification task, the evaluation metrics include accuracy, precision, recall, and F1 score for analyzing the proposed method’s P300 detection capacity. The proposed model performs 69.7\% accuracy for subject A and 79.9\% accuracy for subject B.


\subsection{Character classification estimation}
As shown by Table~\ref{tab:simu}, the accuracy is above the true accuracy that was found for the different number of repetitions, suggesting dependencies across letters and an influence of the user interface.

\subsection{Character classification rate}


The character recognition performance of the weighted ensemble has been shown in Table~\ref{Table V}. According to this table, the proposed method achieves a character recognition accuracy of 98\% in 15 trials for subject A and an accuracy of 91\% for subject B. The ensemble achieves a mean accuracy of 76.5 \% for 5 trials, 87.5 \% accuracy for 10 trials, and 94.5 \% accuracy for 15 trials. Further, we compare the state-of-the-art approaches with ours for 5, 10, and 15 repetitions, as shown in Table~\ref{Table IX}. The proposed method performs better after 5 repetitions \cite{{4454051},{5109302}, {5492691},{7449163}, {doi:10.1080/03772063.2017.1355271}, {9031908}, {9516980},{10018278}} and comparable after 15 repetitions.





\subsection{Ablation study}

A comparison between the character recognition performance of an SPSQ CNN and the weighted ensemble of an SPSQ CNN is presented in Table ~\ref{Table VII}. The  SPSQ CNN shows a character recognition performance of 62\%, 78\%, and 86\% in 5, 10, and 15 trials respectively whereas the weighted ensemble of SPSQ CNN shows a performance of 76.5\%, 87.5\%, and 94.5\% in 5, 10 and 15 trials respectively. These numbers show that the weighted ensemble has a much higher character recognition rate both in low as well as high numbers of repetition of character than a single SPSQ CNN model. The reason for this is that the weighted ensemble can overcome the variance and bias caused by the different models due to being trained on different subsets of the dataset by averaging the scores provided by the base classifiers.

\section{Discussion and Conclusion}
\label{sec:discussion}

Single-trial detection in the EEG signal is a difficult problem as the signal has a low SNR, the number of examples for training is limited, and it is necessary to handle class imbalance in the classification task. Due to the nature of the P300 speller paradigm, for each repetition of a character, there are two target and 10 Non-target signals. While many CNN architectures have been proposed in the literature, most of them can show good performance with multiple repetitions of character.  This work proposes a novel weighted ensemble SPSQ-CNN for the detection of event-related potentials in a lower number of repetitions of character. The method employs a weighted ensemble of SPSQ-CNNs to extract spatiotemporal features from pre-processed signals, enhancing classification performance and reducing classifier variability. The proposed approach achieves 69.7\% accuracy for subject A and 79.9\% for subject B in the P300 classification.

\begin{table*}[]
\centering
\caption{Comparison of average inference time of a single SPSQ-CNN and WE-SPSQ -CNN.}
\label{Table VIII}
\begin{tabular}{@{}c|ccccccccccccccc@{}}
\toprule
\multirow{2}{*}{Method} & \multicolumn{15}{c}{Number of repetitions (epochs/trials)} \\
 & 1 & 2 & 3 & 4 & 5 & 6 & 7 & 8 & 9 & 10 & 11 & 12 & 13 & 14 & 15 \\ \midrule
SPSQ CNN & 2.40 & 3.07 & 3.72 & 4.22 & 4.92 & 5.57 & 6.12 & 6.80 & 7.49 & 8.03 & 8.88 & 9.48 & 9.98 & 12.30 & 13.17 \\
WE-SPSQ CNN & 8.14 & 8.95 & 9.67 & 10.14 & 11.30 & 12.18 & 13.23 & 13.80 & 15.16 & 17.75 & 21.86 & 19.11 & 19.62 & 19.09 & 20.24 \\ \bottomrule
\end{tabular}
\end{table*}


This work implements score-based averaging to improve character recognition accuracy, achieving results ranging from 76.5\% in five repetitions to 94.5\% in fifteen repetitions. The method demonstrates competitive performance, with accuracy reaching 76.5\%, 87.5\%, and 94.5\% for 5, 10, and 15 respectively. The AUC-ROC values for the respective subjects are 69.041 and 72.082. An estimation for the character recognition accuracies in multiple repetitions has been made using these AUC values and has been reported in Table II. The simulated values are much higher than the ones observed. This shows that the character recognition performance is not solely dependent on the P300 prediction performance. 

However, exploring the aggregation of multiple architectures could lead to further advancements in real-time BCI applications. The model does not show consistent improvement with trials but instead recognizes the same number of characters in different epochs. Ensemble models require inference from multiple models and hence take more inference time. This can be improved by using a quantization approach. The performance of the model was analyzed on a dataset collected using 64 different electrodes. In the future, more studies can be done with a lower number of electrodes.

Although the usage of ensemble model for this task shows good performance in terms of accuracy, a significant downside of using emsemble models is that they have a longer inference time as the predictions are being made by multiple models. This is evident from the comparison shown in Table~\ref{Table VIII}. Future works will include data augmentation and the model evaluation across participants as well as experimentation using quantization and quantization aware model training to reduce the inference time required by the model. Another way to deal with the larger inference times is to make concurrent predictions using multithreading. This solution will be explored as well in the future works. It is worthwhile to consider systems that do not require a calibration session, and where a model from one or more users can be used for new users. 

\subsection*{Acknowledgment}
This study was supported by 
IP/IITGN/CSE/YM/2324/05 grants.

\bibliographystyle{unsrt}
\bibliography{mybib.bib}

\end{document}